\documentclass[onecolumn,usenatbib]{mn2e}
\def\lcdm{$\Lambda{\rm CDM}$ }
\def\mnras{MNRAS}
\def\apj{ApJ}
\def\apjl{ApJ Letters}
\def\physrep{Phys Reps}
\def\aj{A J}
\def\aap{A\&A}
\def\nat{Nature}
\def\x{{\bf x}}
 
\def\g{\gamma}
\def\th{\Theta}
\usepackage{epsfig}
\usepackage{graphics}
\author[Yadav, Bharadwaj, Pandey and Seshadri]{Jaswant
  Yadav$^1$\thanks{Email: jaswant@physics.du.ac.in},  
 Somnath Bharadwaj$^2$\thanks{Email: somnathb@iitkgp.ac.in}, 
Biswajit Pandey$^2$\thanks{Email: pandey@cts.iitkgp.ernet.in} and 
T.R.Seshadri$^1$\thanks{Email: trs@physics.du.ac.in}\\
$^1$ Department of Physics \&  Astrophysics, University of Delhi, Delhi
110007,India.\\ 
 $^2$  Department of Physics and Metereology and Centre for
Theoretical Studies,  IIT Kharagpur, 721 302, India. }
\title[Testing homogeneity in SDSS-DR1]{Testing homogeneity on large scales in
  the Sloan Digital Sky Survey Data Release One}   
\begin{document}
\maketitle
\begin{abstract}
The assumption that the universe is homogeneous and isotropic on large
scales is one of the fundamental postulates of cosmology. We have
tested the large scale homogeneity of the galaxy distribution in the  
Sloan Digital Sky Survey Data Release One (SDSS-DR1) using 
 volume limited subsamples extracted from the two equatorial strips
which  are nearly two dimensional (2D).  
The galaxy distribution was projected  on the equatorial plane and
we carried  out a 2D multi-fractal analysis  by counting the number of galaxies 
inside circles of different radii $r$ in the range $5 \, h^{-1} {\rm Mpc}$ to $150 \,
h^{-1} {\rm Mpc}$  centred on galaxies. Different moments of the
count-in-cells were analysed to identify a range of length-scales
($60-70 \, h^{-1} {\rm Mpc}$ to $150 h^{-1} {\rm Mpc}$ ) where the moments
show a power law scaling behaviour and to determine the scaling
exponent which gives the spectrum of generalised dimension $D_q$. 
If the galaxy distribution is homogeneous, $D_q$ does not vary with $q$ and
is equal to the Euclidean dimension which in our case is 2.
We find that $D_q$ varies in the range
$1.7$ to $2.2$. We also constructed mock data from random, homogeneous
point
distributions and from \lcdm N-body simulations with bias $b=1, 1.6$
and $2$,  and analysed these in exactly the same way. The values of
$D_q$ in the random distribution and the unbiased simulations show
much smaller variations  and these are not consistent with the actual
data.  The biased simulations, however, show larger variations in $D_q$ and
these are  consistent with both the random and the actual data.
Interpreting the actual data as a realisation of a biased \lcdm
universe, we conclude that the galaxy distribution is homogeneous on
scales larger than  $60-70 \, h^{-1} {\rm Mpc}$. 
\end {abstract}
\begin{keywords}
 methods: numerical - galaxies: statistics -
cosmology: theory - cosmology: large scale structure of universe
\end {keywords}
\section{Introduction}
The primary aim and objective of all  galaxy
redshift surveys is to determine the large scale structures 
in the universe. Though the galaxy distribution  exhibits a large  
variety of structures starting from groups and clusters, extending to
superclusters  and an interconnected network of filaments which appears to
extend across the whole  universe, we expect the galaxy distribution
to be homogeneous on large scales. 
The
assumption  that the universe is
homogeneous and isotropic on large scales is known as the
Cosmological Principle and this is one of the fundamental  pillars of
cosmology. In addition to determining the large scale structures,
galaxy redshift surveys can also be used to verify that the galaxy
distribution does indeed become homogeneous on large scales and
thereby  validate  the Cosmological Principal. 
Further, these can be used to investigate the scales at which this
transition to homogeneity takes place.
In this paper we
test whether the galaxy distribution in the SDSS-DR1 (\citealt{abaz})  is
{\em actually}   homogeneous on large scales. 

A large variety  of methods have been developed and used to quantify 
the galaxy distribution in redshift surveys, prominent among these
being the two-point correlation function $\xi(r)$ (\citealt{pee}) and
its Fourier transform 
the power spectrum $P(k)$. There now exist very precise estimates of
$\xi(r)$ (eg. SDSS, \citealt{zevi}; 2dFGRS, \citealt{haw})  and the
power spectrum  $P(k)$ (eg.   2dFGRS,  \citealt{perci}; SDSS
\citealt{teg2}) determined from different large redshift surveys.  
On small scales the two point correlation function is  found to be
well described by  the form  
\begin{equation}  
  \xi(r)=(\frac{r}{r_0})^{\g}
\label{eq:1}
\end {equation}
where  $\g =1.75\pm 0.03 $ and $ r_0=6.1\pm 0.2\, h^{-1} {\rm
  Mpc}$ for the SDSS (\citealt{zevi}) and $\g =1.67\pm 0.03
  $ and $ r_0=5.05\pm 0.26\, h^{-1} {\rm Mpc}$ for the 2dFGRS 
(\citealt{haw}). 
The power law behaviour 
of $\xi(r)$  suggests a scale invariant clustering pattern which would
violate homogeneity if
this power-law behaviour
were to extend to arbitrarily large
length-scales. Reassuringly, the power law form for $\xi(r)$ 
does not hold on large scales and it breaks down at $r> 16 h^{-1} {\rm Mpc} $ 
for SDSS and at $r> 20 h^{-1} {\rm Mpc} $ for 2dFGRS. The fact that the values
of $\xi(r)$ fall sufficiently with
increasing $r$ is consistent with the galaxy distribution being homogeneous on 
large
scales. A point to note is that though the $\xi(r)$ determined from
redshift surveys is consistent with the universe being homogeneous at
large scales it does not actually test this. This is because the way
in which $\xi(r)$ is defined and determined from observations refers
to the mean number density of galaxies  and therefore it presupposes
that the galaxy distribution is homogeneous on large scales. 
Further, the mean density which we compute is only that on the scale
of the survey. It will be equal to the mean density in the universe only
if the transition to homogeneity occurs well within the survey region.
To verify
the large scale homogeneity of the galaxy distribution it is necessary
to consider a statistical test which does not presuppose the premise
which is being tested. Here we consider one such test, the
``multi-fractal dimension'' and apply it to the SDSS-DR1. 

 A fractal is a geometric object such that each part of 
it is a reduced version of the whole i.e. it has the same appearance on
all scales. Fractals have been invoked to describe 
many physical phenomena which exhibit self-similarity. A
multi-fractal is an extension of the concept of a fractal. It
incorporates the possibility that the particle distribution in
different density environments may exhibit a different scaling or
self-similar behaviour. 
The fact that the galaxy clustering  is scale-invariant  over a  range
of 
length-scales led \citet{peter} to propose that the galaxies had a
fractal distribution. The later analysis of \citet{coleman} seemed to
bear out such a proposition whereas \citet{bor} claimed that 
the fractal description was valid only on small scales and the 
galaxy distribution was consistent with homogeneity on large scales. 
A purely fractal distribution 
would not be homogeneous on any length-scale and this would violate the
Cosmological Principle. Further, 
the mean density would decrease if it were to be
evaluated for progressively  larger volumes and this would manifest
itself as an increase in the correlation length $r_0$ (eq. \ref{eq:1})
with the size of the sample. However, this simple prediction of the
fractal interpretation is not  
supported by data, instead $r_0$ remains constant for volume limited samples
of CfA2 redshift survey with increasing depth \citep{martinez}.

The analysis of the ESO slice project \citep{gujo} confirms large scale
 homogeneity whereas  the  analysis  of volume limited samples of
 SSRS2 \citep{cappi} is consistent with both the   scenarios of
 fractality and homogeneity. A similar analysis \citep{haton} carried
 out on  APM-Stromlo survey exhibits a fractal behaviour with a
 fractal dimension of $ 2.1\pm 0.1$ on scales up to $40\, h^{-1}{\rm
 Mpc}$. Coming to the fractal analysis of the LCRS, \citet{amen} 
find a fractal behaviour on scales less than $\sim$ 
$30 h^{-1}\,{\rm Mpc}$ but are inconclusive about the transition to
homogeneity. A multi-fractal analysis by \citet{bharad1} shows that
 the   LCRS exhibits homogeneity on the scales $80$ to $200 \,
 h^{-1}\, {\rm Mpc}$. 
The analysis of \citet{kur}  shows
this to occur at a length-scale of $\sim 30 \, h^{-1}\,{\rm Mpc}$, 
whereas \citet{best} fails to find a transition to homogeneity even on
the largest scale analysed. The fractal analysis of the 
PSCz  \citep{pan} shows   a transition to homogeneity on  scales
 of $30\, h^{-1} {\rm Mpc}$.   Recently \citet{baris} have  performed a
 fractal analysis of SDSS EDR and find that a  fractal distribution
 continues to  length-scales of $200\,h^{-1}{\rm Mpc}$ whereas  
\citet{hog} analyse the  SDSS LRG  to find 
a convergence to homogeneity at a scale of  around $70\, h^{-1} {\rm
 Mpc}$. 

In this paper we use the multi-fractal analysis to study the scaling
properties of the galaxy distribution in the  SDSS-DR1 and test if it
is consistent with homogeneity on large scales. The SDSS  is the
largest galaxy survey available at present. For the current analysis
we have used  volume limited subsamples   extracted from the two equatorial
strips of the SDSS-DR1. This reduces the number of galaxies but offers
several advantages. The variation in the number density in these
samples are independent of the details of the luminosity function and
is caused by the clustering only. The larger area and depth of these
samples provide us the scope  to investigate the scale of homogeneity
in greater detail.

The \lcdm  model with $\Omega_{m0}=0.3$, $\Omega_{\Lambda0}=0.7$,
$h=0.7$ and a  featureless, adiabatic,  scale invariant
primordial power spectrum is currently believed to be the 
minimal model which is consistent with most 
cosmological data (\citealt{efst}; \citealt{perci1}; \citealt{teg1}). 
Estimates of the two point correlation function $\xi(r)$ 
(LCRS, \citealt{tuck}; SDSS, \citealt{zevi}; 2dFGRS, \citealt{haw})  
and the power spectrum  $P(k)$ (LCRS, \citealt{lin};  2dFGRS,
 \citealt{perci}; SDSS,  \citealt{teg2}) are all consistent with this
 model.  In this paper we use N-body simulations to determine the
 length-scale where the transition to homogeneity occurs in the \lcdm
 model and test if the actual data is consistent with this.

 There are various other probes which  test the cosmological
 principle. The fact that  the Cosmic Microwave Background Radiation (CMBR) is
 nearly isotropic $(\Delta T/T \sim  10^{-5})$ can be used to infer  that our
 space-time is locally  very well described by the Friedmann-Robertson-Walker
 metric (\citealt{ehl}). Further, the 
 CMBR  anisotropy  at large angular scales $(\sim 10^{o})$ constrains
 the  {\it rms}  density fluctuations to $\delta\rho/\rho
 \sim 10^{-4} $ on  length-scales of $1000\,h^{-1}{\rm Mpc}$
 (e.g. \citealt{wu}). The analysis of deep radio 
surveys (e.g. FIRST, \citealt{bale}) suggests the distribution to be
 nearly isotropic on large scales. By comparing the predicted 
multipoles of the X-ray Background to those observed by HEAO1
 (\citealt{sch}) the   
fluctuations in amplitude are found to be consistent with the homogeneous 
universe (\citealt{lah}). The absence of big  voids in the
distribution of Lyman-$\alpha$ absorbers is
inconsistent with a fractal model (\citealt{nus}).

A brief outline of the paper follows. In Section 2 we describe the
data and the method of analysis, and Section 3 contains results and
conclusions. 
\section{Data and method of analysis}
\subsection{SDSS and the N-body data}

\begin{figure}
\rotatebox{-90}{\scalebox{.6}{\includegraphics{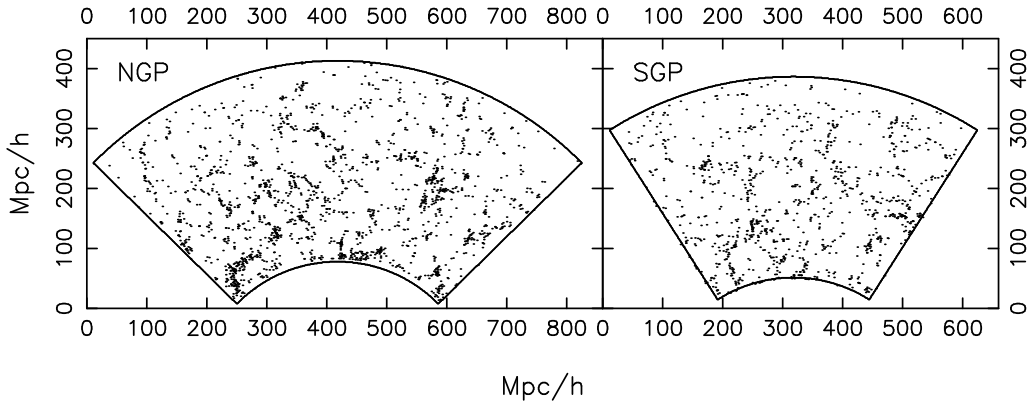}}}
\caption{This shows the two dimensional galaxy distribution in in  the NGP
  and SGP subsamples that have been analysed here.} 
\label{fig:1}
\end{figure}

 SDSS is the largest  redshift survey at present and 
 our analysis is based on the publicly available SDSS-DR1
 data (\citealt{abaz}). 
Our analysis is limited to the two equatorial strips 
 which are centred along the celestial equator ($\delta=0^{\circ}$),
 one  in the Northern Galactic  Cap (NGP)  spanning  $91^{\circ}$   in
 {\it r.a.} and the other  Southern Galactic Cap (SGP) spanning
 $65^{\circ}$ in  {\it r.a.},  their thickness varying 
within $\mid \delta \mid \le 2.5^{\circ}$ in {\it dec.}
We  constructed  volume limited subsamples extending from  
 $z=0.08$ to $0.2$ in redshift ({\it i.e.} $235\,h^{-1}
 {\rm Mpc}$ to  $571\,h^{-1}{\rm Mpc}$ comoving  in the  radial
 direction) by restricting  the absolute 
 magnitude   range to  $-22.6\leq M_r  \leq -21.6$.   
The resulting subsamples are two thin wedges of varying thickness aligned
 with the equatorial plane. Our analysis is restricted to slices of
 uniform thickness  $\pm 4.1 \,h^{-1}{\rm Mpc}$ along  the 
 equatorial plane extracted out of the wedge shaped regions. These
 slices are 
 nearly 2D with the radial extent and the
 extent along {\it r.a.} being much 
 larger than the thickness. We have projected the galaxy distribution
 on the equatorial plane and analysed the resulting 2D distribution
 (Figure \ref{fig:1}).   
  The SDSS-DR1 subsamples that we
 analyse here contains  a  total of 3032 galaxies and the subsamples
are  exactly same as those  analysed in \citet{pandey}.  
 We have used a Particle-Mesh (PM) N-body code to simulate the dark
 matter distribution at the mean redshift  $z=0.14$ of our
subsample. A comoving volume of $[645  h^{-1} {\rm Mpc}]^3$ is
 simulated using $256^3$ particles on a
$512^3$ mesh with grid spacing $1.26  h^{-1} {\rm Mpc}$. The set of values
$(\Omega_{m0},\Omega_{\Lambda0},h)=(0.3,0.7,0.7)$ were used for the
cosmological parameters, and we used  a \lcdm power
spectrum characterised by a spectral index $n_s=1$ at large-scales and
with a value $\Gamma=0.2$  for the shape parameter.The power spectrum
was normalised to $\sigma_8=0.84$ (WMAP, \citealt{sperg})
. Theoretical considerations and simulations suggest that galaxies may
 be biased tracer of the underlying dark matter distribution (e.g., 
 \citealt{kais}; \citealt{mo}; \citealt{dekel}; \citealt{taru} and  
 \citealt{yoshi}).
A  ``sharp cutoff'' biasing scheme (\citealt{cole}) was used to
generate particle distributions. This is a local biasing scheme where the 
probability of a particle being
selected as a galaxy is a function of local density only.
In this scheme the final dark-matter distribution generated by the
 N-body simulation was first smoothed with a Gaussian of width $5
 h^{-1} {\rm Mpc}$. Only the particles which lie in regions where the
 density contrast exceeds a critical value were selected as galaxy.
 The values of the critical density contrast were chosen so as to
 produce particle distributions with a low bias $b=1.2$ and a high
 bias $b=1.6$.
 An observer is placed at a suitable location inside
 the N-body simulation cube and we use the peculiar velocities to
 determine the particle positions in redshift space.
 Exactly the same number of   particles distributed over  the
 same volume as the actual data was extracted from the simulations
 to produce simulated NGP and SGP slices. The simulated slices were
 analysed in exactly the same way as the actual data. 

\subsection{Methods of Analysis}
A  fractal point distribution is usually characterised 
 in terms of its fractal dimension. There are different ways to
 calculate this, and the correlation dimension is one of the methods
 which is   of  particular relevance to the analysis of galaxy
 distributions.  The formal definition of the correlation dimension
 involves a limit  which is meaningful only when the number of
 particles is infinite and  hence this cannot  be applied to  galaxy
 surveys with a limited number of galaxies. To overcome this we adopt 
a  ``working  definition'' which can be applied to a finite 
 distribution of $N$ galaxies. It should be noted that our
 galaxy distribution is effectively two dimensional, and we have
largely restricted our discussion to this situation. 

 Labelling the galaxies from $1$ to $N$,
 and using $\x_i$ and $\x_j$ to denote the comoving coordinates of the
 $i$~th and the  $j$~th  galaxies respectively, the number of
 galaxies within a circle of comoving radius $r$ 
 centred on the $i$~th galaxy is 
\begin{equation}
  n_i(r)=\sum_{j=1}^{N}\th(r-\mid \x_i-\x_j \mid)
\label{eq:2}
\end{equation}
where  $\th(x)$ is the  Heaviside function defined such that
 $\th(x)=0$ for $x<0$ and $\th(x)=1$ for $x\ge0$. Averaging $n_i(r)$
by choosing  $M$ different galaxies as centres and dividing by the
 total number of galaxies gives us 
\begin{equation} 
  C_2(r)=\frac{1}{MN}\sum_{i=1}^{M}n_i(r)
\end{equation}
which may be interpreted as the probability of finding a galaxy within
a circle of radius $r$ centred on another galaxy. 
If  $C_2(r)$ exhibits a power law scaling relation $C_2(r)\propto
r^{D_2}$, the exponent  $D_2$ is defined to be the correlation
dimension. Typically, a power law scaling relation will hold only
over a limited range of length-scales $r_1 \le r \le r_2$, and it may
so happen that the galaxy distribution has different correlation
dimensions over different ranges of length-scales. 

It is  clear  that $C_2(r)$ is closely related to the volume
integral of the  two point correlation function  $\xi(r)$. In a situation
where this has  a power law behaviour $\xi(r)=(\frac{r}{r_0})^{\g}$, 
the correlation dimension is $D_2=2-\g$ on scales
$r<r_0$. Further, we expect $D_2=2$ on large scales where the galaxy
distribution is expected to be homogeneous and isotropic. 

In the usual analysis  the two point correlation does not fully
characterise all the statistical properties of the galaxy
distribution, and it is necessary to also consider the higher order
correlations {\it eg.} the three point and higher
correlations. Similarly, the full statistical  quantification  of  a
fractal  distribution also requires  a hierarchy of scaling indices. 
The multi-fractal analysis used  here does 
 exactly this. It provides a continuous spectrum of
generalised dimension $D_q$,  the Minkowski-Bouligand dimension,
which is defined for a range of $q$. 

 The definition of  the generalised dimension  $D_q$ closely follows
 the definition of the correlation dimension $D_2$, the only difference
 being that we use the $(q-1)$th moment of $n_i(r)$. The quantity
 $C_2(r)$ is now generalised to 
\begin{equation} 
  C_q(r)=\frac{1}{MN}\sum_{i=1}^{M}[n_i(<r)]^{q-1}
  \label{eq:3}
\end{equation}
which is used to define the Minkowski-Bouligand dimension 
\begin{equation} 
D_q=\frac{1}{q-1}\frac{d\log{C_q(r)}}{d\log{r}}
\label{eq:4}
\end{equation}
Typically $C_q(r)$ will not exhibit the same scaling behaviour over
the entire range of length-scales, and it is possible that the 
spectrum of  generalised dimension will be different  in different
ranges of length-scales. 
The correlation dimension corresponds to the generalised dimension
at $q=2$, whereas $D_1$  corresponds to  the box counting dimension. 
The other integer values of $q$ are related to the scaling of higher order
correlation functions.
A mono-fractal is characterised by a single scaling
exponent {\it ie.} $D_q$ is a constant independent of $q$, whereas the
full spectrum of generalised dimensions is needed to characterise a
multi-fractal.   The positive values of $q$ give more weightage to the
regions with high number  density whereas the  negative values of $q$
give more weightage to the underdense  regions. Thus we may interpret 
$D_q$  for  $q > 0$ as characterising the scaling behaviour of the
galaxy distribution in the high density regions like clusters whereas
$q<0$ characterises the scaling inside voids. In the situation where
the galaxy distribution  is homogeneous and isotropic on large scales,
we expect  $D_q=2$ independent of the value of $q$.

There are a variety of different algorithms which can be used  to
calculate the generalised  dimension, the Nearest Neighbour
Interaction(\citealt{badi}) and the Minimal Spanning Tree
(\citealt{suth}) being some of them. We have used the correlation
integral method which we present below. 

The two subsamples, NGP and SGP contain 1936 and 1096 galaxies
respectively and they were analysed separately. For each galaxy in the
subsample we considered a circle of  radius $r$ centred on the
galaxy and counted the number of other galaxies within the circle to
determine  $n_i(r)$ (eq. \ref{eq:2}). The radius $r$ was increased
starting from $5\, h^{-1} \, {\rm Mpc}$ to the largest value where the
circle lies entirely within the subsample boundaries. 
The values of $n_i(r)$ determined using different galaxies as
centres were then averaged to determine $C_q(r)$ (eq. \ref{eq:4}). It
should be noted 
that the number of centres falls with increasing $r$, and for the NGP
there   are $\sim 800$ centres for $r=80h^{-1} \, {\rm  Mpc}$  with the value
falling  to $\sim 100$ for a radius of $r=150h^{-1} \, {\rm
  Mpc}$. The large scale behaviour of $C_q(r)$ was analysed to
determine the range of length-scales where it  exhibits a scaling
behaviour  and to identify the scaling exponent $D_q$ as a function 
of $q$.

In addition to the actual data,  we have also constructed and analysed
random distributions of points. The random data  contains
exactly the same number points as there are galaxies in the 
actual data distributed over exactly the same region as the actual
NGP and SGP slices. The random data  are homogeneous and
isotropic  by construction, and  the results of the multi-fractal
analysis of this data gives definite predictions for the
results expected if the galaxy distribution were homogeneous and
isotropic.  The random data and the simulated slices extracted from
the N-body simulations were all analysed in exactly the same way as
the actual data. We have used 18 independent realisations of the
random  and simulated slices to estimate the mean and the
$1-\sigma$ error-bars of the spectrum of generalised dimensions
$D_q$. 
\section{Results and Discussions}

\begin{figure}
\rotatebox{-90}{\scalebox{.6}{\includegraphics{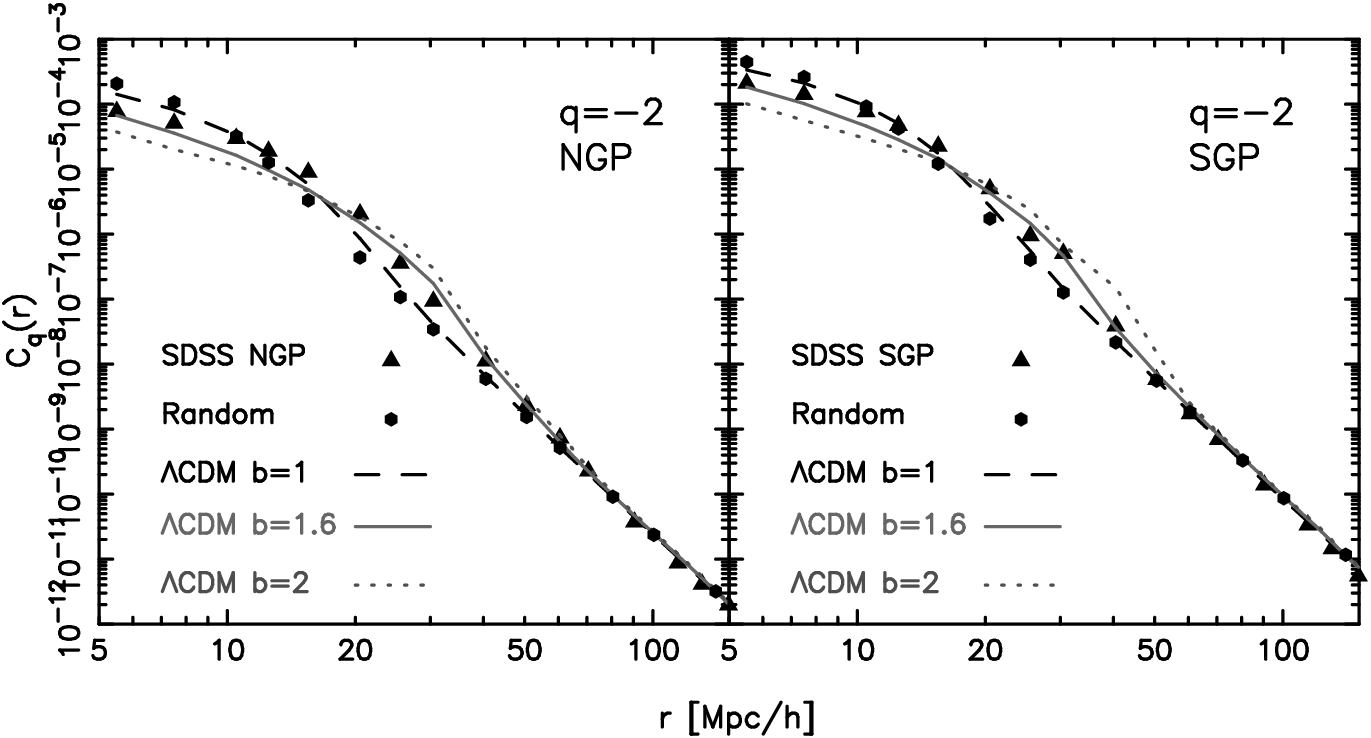}}}
\caption{This shows $C_q(r)$ at $q=-2$ for the actual data, the random
data and the simulated slices.}
\label{fig:f1}
\end{figure}

\begin{figure}
\rotatebox{-90}{\scalebox{.6}{\includegraphics{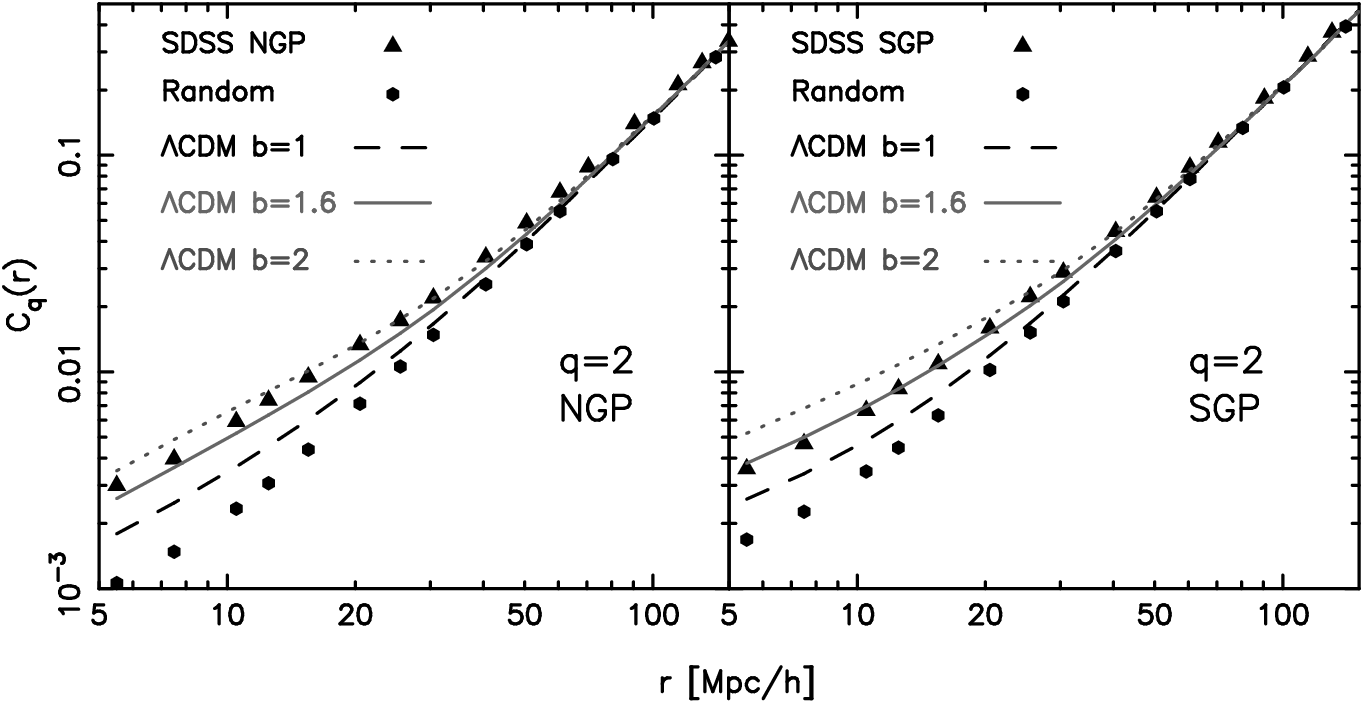}}}
\caption{This shows $C_q(r)$ at $q=2$ for the actual data, the random
data and the simulated slices.}
\label{fig:f2}
\end{figure}
\begin{figure}
\rotatebox{-90}{\scalebox{.6}{\includegraphics{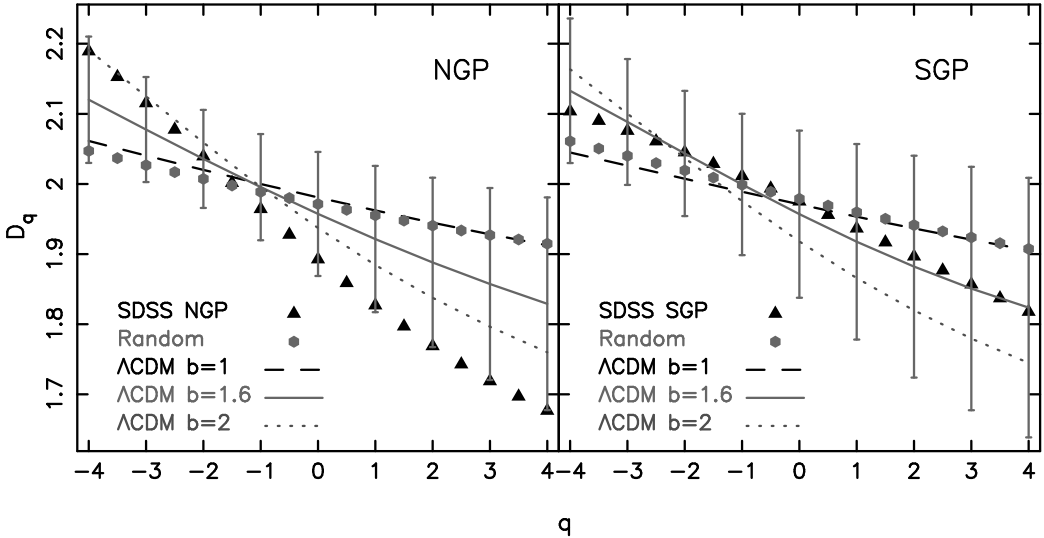}}}
\caption{This shows the spectrum of generalised dimensions $D_q$ as a
function of $q$ for the actual data, the random
data and the simulated slices on length scales of from
$60-70 \, h^{-1} \, {\rm Mpc}$ to $150 \, h^{-1} \, {\rm Mpc}$.
The error bars shown are for \lcdm model with bias=1.6.}
\label{fig:f3}
\end{figure}

Figures \ref{fig:f1} and \ref{fig:f2} show $C_q(r)$  at $q=-2$
 and $2$, respectively,  for the actual data, for one realisation of
 the random 
slices and for one realisation of the  simulated slices for each value
 of the bias.   The behaviour of $C_q(r)$ at other 
 values of $q$ is similar to the ones shown here. Our analysis is
 restricted to $-4 \le q \le 4$. 
We find that $C_q(r)$ does not exhibit a scaling
behaviour at small scales $(5\, h^{-1} \, {\rm Mpc} \le r \le 40 \,
h^{-1} \, {\rm Mpc})$. Further, the small-scale  behaviour of $C_q(r)$
in the actual data is different from that of the random slices and is
roughly consistent with the simulated slices for $b=1.6$.  We find
 that $C_q(r)$ shows a scaling behaviour on length-scales of 
 from somewhere around $60-70 \, h^{-1} \, {\rm Mpc}$ to $150 \,
 h^{-1} \, {\rm Mpc}$.  Although the actual data, the random and
 simulated slices all appear to converge over this range of
 length-scales indicating that they are all roughly consistent with
 homogeneity, there are small differences in the slopes. We have used
 a least-square fit to determine the scaling exponent or
 generalised dimension $D_q$ shown in Figure \ref{fig:f3}. 

Ideally we would expect $D_q=2$ for a  two dimensional   
homogeneous and isotropic  distribution. 
We find that for the actual data $D_q$  varies in the range $1.7$ to
$2.2$ in the NGP and $1.8$ to $2.1$ in the SGP on large-scales. In both the slices the
value of $D_q$  decreases with increasing $q$, and it crosses $D_q=2$
somewhere around $q=-1$. The variation of $D_q$ with $q$ shows a
similar behaviour in the random slices, but the range of variation is
much smaller $(1.9 \le D_q \le 2.1)$.  Comparing the actual data with
the random data we find that the actual data lies outside the
$1-\sigma$ error-bars of the random data (not shown here) for most of
the range of $q$ except around $q=-1$ where $D_q=2$ for
both the actual and random data. Accepting this at face value would
imply that the actual data is not homogeneous at large scales. 

Considering the simulated data, we find that the variation in $D_q$
depends on the value of the bias $b$. For the unbiased simulations
$D_q$ shows very small variations $(1.9 \le D_q \le 2.1)$ and the
results are very close to those of the random data. We find that
increasing the bias causes the variations in $D_q$ to increase. In all
cases $D_q$ decreases with increasing $q$ and it crosses $D_q=2$
around $q=-1$. Increasing the bias has another effect in that it
results in larger $1-\sigma$ error-bars. 

Comparing the simulated data with the random data and the actual data
we find that the unbiased simulations are consistent with the random
data but not the actual data. The actual data lies outside the
$1-\sigma$ error-bars of the unbiased \lcdm model. 
This implies that the unbiased \lcdm
model has a transition to homogeneity at $60-70 \, h^{-1} \, {\rm
  Mpc}$.  The spectrum of generalised dimensions as determined from
the unbiased simulations on
length-scales $60-70 \, h^{-1} \, {\rm  Mpc}$ to $150 \, h^{-1} \, {\rm
  Mpc}$ is different from that of the actual data {\it ie.} the
unbiased \lcdm model fails to reproduce the large scale   properties
of the galaxy distribution in our volume limited subsamples of the
SDSS-DR1.  

 The simulations with bias $b=1.6$ and $b=2$ have larger $1-\sigma$
 error-bars and these are consistent with both the random and the
 actual data.  Interpreting the actual data as being a realisation of
 a biased \lcdm universe, we conclude that it has a transition to
 homogeneity at $60-70 \, h^{-1} \, {\rm  Mpc}$ and the galaxy
 distribution is homogeneous on scales larger than this.

The galaxy subsample analysed here contains the most  luminous
galaxies in the SDSS-DR1. Various investigations have shown the bias
to increase with luminosity (\citealt{nor}; \citealt{zevi}) and the
subsample analysed here is  expected to be 
biased with respect to the underlying dark matter distribution.  
\citet{sel} have used the halo model in conjunction with weak lensing
to determine the bias for a number of subsamples with different
absolute magnitude ranges. The brightest sample which they have
analysed has galaxies with absolute magnitudes in the range $-23 \le
M_r \le -22$ for which they find a bias $b=1.94 \pm 0.2$. Our results
are consistent with these findings. 

A point to note is that the $1-\sigma$ error-bars of the spectrum of
generalised dimension $D_q$ increases with the bias. This can be
understood in terms of the fact that  $C_q(r)$ is related to  volume 
integrals of the correlation functions which receives contribution from
all length-scales. The fluctuations in $C_q(r)$ can also be related to
volume integrals of the correlation functions. Increasing the bias
increases the correlations on small scales $(\le 40-50 \, h^{-1} {\rm
  Mpc})$  which contributes to the fluctuations in $C_q(r)$ at large 
scales and causes the fluctuations in $D_q$ to increase. 

The galaxies in nearly all redshift surveys appear to be distributed
along filaments. These filaments appear to be interconnected and they
form a complicated network often referred to as the ``cosmic web''. 
These filaments are possibly the largest coherent structures in
galaxy redshift surveys. Recent analysis of volume limited subsamples
of the LCRS \citep{bharad2} and the same SDSS-DR1 subsamples analysed
here \citep{pandey} shows the filaments to be statistically
significant features of the galaxy distribution on  length-scales $\le
70-80 \, h^{-1} {\rm   Mpc}$ and not beyond. Larger filaments present
in the galaxy distribution are not statistically significant and are
the result of chance alignments. Our finding that the galaxy
distribution is homogeneous on scales larger than $60-70 \, h^{-1}
{\rm   Mpc}$ is consistent with the size of the largest statistically
significant coherent structures namely the filaments. 
\section*{Acknowledgements}
SB would like to acknowledge financial support from the Govt. of
India, Department of Science and Technology (SP/S2/K-05/2001).
JY and BP  are  supported by  fellowships of the
Council of Scientific and Industrial Research (CSIR), India.
JY and TRS would like to thank 
IUCAA for the facilities at IUCAA Reference Centre at Delhi
University. TRS thanks IUCAA for the support provided through the
Associateship Program. 

The SDSS-DR1 data was
downloaded from the SDSS skyserver http://skyserver.sdss.org/dr1/en/.
    Funding for the creation and distribution of the SDSS Archive has been
provided by the Alfred P. Sloan Foundation, the Participating
Institutions, the National Aeronautics and Space Administration, the
National Science Foundation, the U.S. Department of Energy, the Japanese
Monbukagakusho, and the Max Planck Society. The SDSS Web site is
http://www.sdss.org/.

    The SDSS is managed by the Astrophysical Research Consortium (ARC) for
the Participating Institutions. The Participating Institutions are The
University of Chicago, Fermilab, the Institute for Advanced Study, the
Japan Participation Group, The Johns Hopkins University, the Korean
Scientist Group, Los Alamos National Laboratory, the Max-Planck-Institute
for Astronomy (MPIA), the Max-Planck-Institute for Astrophysics (MPA), New
Mexico State University, University of Pittsburgh, Princeton University,
the United States Naval Observatory, and the University of Washington.

\end{document}